\documentclass[12pt,preprint]{aastex}
\newcommand{\f}{f^{(\nu)}}
\newcommand{\be}{\begin{equation}}
\newcommand{\ee}{\end{equation}}

\shorttitle{Partial suppression of radial orbit instability}
\shortauthors{Trenti \& Bertin}

\begin{document}

%% LaTeX will automatically break titles if they run longer than
%% one line. However, you may use \\ to force a line break if
%% you desire.

\title{Partial suppression of the radial orbit instability in stellar systems}

%% Use \author, \affil, and the \and command to format
%% author and affiliation information.
%% Note that \email has replaced the old \authoremail command
%% from AASTeX v4.0. You can use \email to mark an email address
%% anywhere in the paper, not just in the front matter.
%% As in the title, use \\ to force line breaks.

\author{M. Trenti\altaffilmark{1}}
\affil{Scuola Normale Superiore, Piazza dei Cavalieri 7, I-56126 Pisa, Italy}
\email{m.trenti@sns.it}

\and

\author{G. Bertin}
\affil{Dipartimento di Fisica, Universit\`{a} di Milano, via
Celoria 16, I-20133 Milano, Italy}
\email{giuseppe.bertin@unimi.it}

%% Notice that each of these authors has alternate affiliations, which
%% are identified by the \altaffilmark after each name.  Specify alternate
%% affiliation information with \altaffiltext, with one command per each
%% affiliation.

\altaffiltext{1}{present address: Space Telescope Science Institute, 3700 San Martin Drive Baltimore MD 21218 USA}

\begin{abstract}
It is well known that the simple criterion proposed originally by
Polyachenko and Shukhman (1981) for the onset of the radial orbit
instability, although being generally a useful tool, faces
significant exceptions both on the side of mildly anisotropic
systems (with some that can be proved to be unstable) and on the
side of strongly anisotropic models (with some that can be shown
to be stable). In this paper we address two issues: Are there
processes of collisionless collapse that can lead to equilibria of
the exceptional type? What is the intrinsic structural property
that is responsible for the sometimes noted exceptional stability
behavior? To clarify these issues, we have performed a series of
simulations of collisionless collapse that start from homogeneous,
highly symmetrized, cold initial conditions and, because of such
special conditions, are characterized by very little mixing. For
these runs, the end-states can be associated with large values of
the global pressure anisotropy parameter up to $2K_r/K_T \approx
2.75$. The highly anisotropic equilibrium states thus constructed
show no significant traces of radial anisotropy in their central
region, with a very sharp transition to a radially anisotropic
envelope occurring well inside the half-mass radius (around
$0.2~r_M$). To check whether the existence of such almost
perfectly isotropic ``nucleus" might be responsible for the
apparent suppression of the radial orbit instability, we could not
resort to equilibrium models with the above characteristics and
with analytically available distribution function; instead, we
studied and confirmed the stability of configurations with those
characteristics by initializing N-body approximate equilibria
(with given density and pressure anisotropy profiles) with the
help of the Jeans equations.
\end{abstract}


%%%%%%%%%%%%%%%%%%%%%%%%%%%%%%%%%%%%
\section{Introduction}

The study of the stability of a system of which the distribution
function is available in analytical form can be carried out along
three different approaches: N-body simulations (e.g., see
\citealt{hen73}; \citealt{mer85b}; \citealt{bar86b}; \citealt{agu90};
\citealt{all90}; \citealt{sti91b}; and following papers), linear modal
analysis
(\citealt{pol81,pal87,pal88,wei89,wei91,wei94,sah91,sah92,ber94}; see
also \citealt{fri84,pal93}, and references therein), and energy
principles (e.g., \citealt{syg84,kan85,goo88}, and references
therein). Spherical stellar systems with too many radial orbits have
thus been found to be unstable to perturbations that break the
spherical symmetry and remove the excess of kinetic energy in the
radial degree of freedom (see \citealt{fri84} and
\citealt{pal93}). The existence of such radial orbit instability was
first investigated by \citet{pol81}, who proposed an ``empirical
criterion" for the onset of instability ($2K_r/K_T > 1.7 \pm 0.25$)
based on the global content of kinetic energy in the radial with
respect to that in the tangential degrees of freedom. Following
investigations \citep{bar85,mer85b,agu90,pal93} noted that different
families of models may exhibit different anisotropy thresholds for the
instability, thus widening the uncertainty interval around the value
of $1.7$ suggested earlier. For example, while for the so-called
$f_{\infty}$ and $\f$ models a threshold value similar to that
suggested by Polyachenko and Shukhman may be applicable (see
\citealt{ber89} and \citealt{tre04}, hereafter TB05), for the family
of \citet{deh93} density profiles with an Osipkov-Merritt type
\citep{osi79,mer85} of anisotropy profile it has been argued that such
value is as high as $\approx 2.5$ \citep{mez97}. On the other hand,
systems with an arbitrarily small content of radial anisotropy can be
unstable, although with very small growth rates (as shown by
\citet{pal87} by means of a linear modal analysis).

The radial orbit instability is thought to play an important role
during the formation of self-gravitating structures from
collisionless collapse via incomplete violent relaxation, a
process that has been argued to be the main driver for the
formation of elliptical galaxies \citep{van82}. In fact, if a
system starts from sufficiently cold initial conditions (i.e. with
a low initial virial ratio $u=(2K/|W|)_{t=0} \lesssim 0.15$), it
collapses with stars falling in almost radially toward the center,
often ending up in triaxial configurations, the origin of which is
attributed to the radial orbit instability
\citep{pal90,udr93,hjo95}. In general, when such collisionless
collapse leads to ``realistic" final configurations, the values of
pressure anisotropy achieved at the end of the collapse turn out
to be consistent with the threshold value associated with the
instability criterion proposed by Polyachenko and Shukhman (1981).
Currently, galaxy formation is approached in the generally
accepted cosmological context of hierarchical clustering
\citep[see e.g.][ and references therein]{mez03}, but the
mechanism of incomplete violent relaxation remains an important
ingredient and thus quantifying the effects of the radial orbit
instability is relevant to the explanation of the observed
flattening of gravitational structures in cosmological simulations
(see \citealt{hop05}).

In this paper we present the results of some numerical experiments that
are aimed at clarifying the following two issues: Are there processes of
collisionless collapse able to lead to equilibria of the exceptional
type, that is equilibria violating the criterion proposed by Polyachenko
and Shukhman (1981)? What is the structural property that makes some
models with relatively low levels of global radial anisotropy unstable
and others with high levels of global radial anisotropy stable?

As to the first issue, in a set of simulations of collisionless
collapse (\citealt{tre04b}, hereafter TBvA05) we have investigated the
role of mixing in phase space on the relaxation of the end products of
the simulations. We have thus run, for comparison, a number of
experiments in which phase mixing is inefficient. These are
simulations with highly symmetric initial conditions. In spite of the
cold initial conditions, with $u < 0.1$, and of the high level of
radial anisotropy achieved in the final configurations, with
$2K_r/K_T$ sometimes above $2.5$ (during the process of collapse
values of $2K_r/K_T$ above $10$ are reached), the radial orbit
instability did not develop. From inspection of the structure of the
final products of these simulations, it appears that most likely the
physical factor at the basis of the observed partial suppression of
the radial orbit instability is the existence of an almost perfectly
isotropic central region that is realized in these systems, which
appears to be more efficient for more concentrated models. This is the
clue that we consider in order to address the second issue raised
above.

In fact, we note that at fixed content of global anisotropy
$2K_r/K_T$, the $\f$ and $f_{\infty}$ models (for which the threshold
value of $2K_r/K_T$ is close to 1.7) are less isotropic locally, in
their core, than similar systems with anisotropy profiles of the type
introduced by \citet{osi79} and \citet{mer85}, as studied by
\citet{mez97}, which in turn are less isotropic in their central
regions than the equilibrium states obtained at the end of our set of
simulations. In the opposite direction, we recall that the generalized
polytropic spheres, with small content of global anisotropy, found to
be unstable by \citet{pal87} are characterized by a constant and
finite local anisotropy level down to $r=0$. To confirm this picture
and to study the effects on the radial orbit instability of central
density and anisotropy profiles decoupled from each other, we cannot
resort to equilibrium models with the above characteristics and with
analytically available distribution function. We have then constructed
with the help of the Jeans equations N-body equilibria with density
and anisotropy profiles qualitatively similar to those found as end
states for the simulations of collisionless collapse violating the
criterion proposed by \citet{pol81} and thus demonstrated that some
configurations with $2K_r/K_T$ up to $\approx 2.9$ do not show
evidence of rapid evolution.

The paper is organized as follows. In the next Section we describe
some special processes of collisionless collapse leading to highly
anisotropic quasi-equilibrium states. In Sect.~\ref{sec:rho_alpha} we
investigate the evolution of candidate collisionless equilibrium
configurations constructed numerically starting from the Jeans
equations and identify a quasi-equilibrium state with $2K_r/K_T
\approx 2.9$. Conclusions are presented in Sect.~\ref{sec:con}. In the
Appendix we outline the method used to generate approximate
equilibrium configurations with given density and anisotropy profiles
starting from the Jeans equations.

%%%%%%%%%%%%%%%%%%%%%%%%%%%%%%%%
%%%%%%%%%%%%%%%%%%%%%%%%%%%%%%%%%%%%%%%%%%%%%%%%%%%%%%%%%%%%%%%%%%%%%%%%%%%%%%%%
\section{Exceptionally stable equilibria resulting from special processes
of collisionless collapse} \label{sec:simulations}

Most of the numerical simulations discussed in this paper have
been carried out with the recently developed particle-mesh code
described in TBvA05 \citep[for further tests and
details see also][]{tren04}. The code solves the Poisson equation by
expanding the density and the potential in spherical harmonics and
is well suited for the study of the stability of collisionless
systems (TB05). We have also run a few simulations with the fast
tree code GyrFalcON \citep{deh00,deh02} to check, with success,
that our results are not biased by the specific choice of the
numerical code.

As briefly noted in the Introduction, it is well known (see
\citealt{van82} and many following papers) that the collapse of a cold
cloud of stars can lead, on a dynamical time-scale, to the formation of
an equilibrium structure characterized by a quasi-isotropic core and a
radially anisotropic halo with projected density profiles similar to the
$R^{1/4}$ law \citep{dev48}. The mechanism of incomplete violent
relaxation \citep{lyn67} has thus been advocated as a possible scenario
for the formation of elliptical galaxies. Several investigations
\citep{van82,mcg84,lon91} have shown that the outcome of the
simulations generally resembles observed systems provided the initial
conditions are clumpy. This condition finds a natural interpretation in
the current cosmological framework for structure formation, as the use
of clumpy initial conditions in collisionless collapse simulations may
effectively mimic some aspects of hierarchical merging. Simulations of
mergers of stellar systems, i.e. of incomplete violent relaxation, are
generally considered as an important tool to study the properties of
elliptical galaxies, not only in view of an explanation of the origin of
the universality of the observed surface brightness profiles, but also
in relation to the origin and evolution of the Fundamental Plane
\citep{nip03,gon03}. Observational evidence is now accumulating in
support of the importance of mergers as a formation mechanism for
elliptical galaxies \citep[e.g. see][]{vanDok03,bel04}, although many
arguments and new findings tend to point to a merger process in which
dissipation plays a significant role.  In addition, \citet{kho03} (see
also \citealt{jes05}) argue from semi-analytic modeling that $50~\%$ of
present days ellipticals are relics of mergers between spheroidal
systems.

For the present study, we focus on a particular set of simulations that
start from homogeneous spherically symmetric configurations obtained by
symmetrizing a clumpy configuration with ten cold clumps,
the kinetic energy of which is in the collective motion of their center
of mass. In most of the runs $u= (2K/|W|)_{t=0} < 0.1$; the clump radius
is taken to be equal to one half of the half-mass radius of the
system. The symmetrization is performed by accepting the radius and the
magnitude of the velocity of each simulation particle, following the
procedure to generate clumpy conditions discussed in TBvA05, and by
redistributing uniformly the angular variables in both position and
velocity space. This initialization procedure leads to a smooth initial
density profile, decreasing approximately linearly in radius ($\rho(0)
\approx 2 \rho(r_M)$), which creates a potential well that is deeper
than the one of a uniform sphere with the same cutoff radius. The
initial system is non-rotating and isotropic. Mild correlations in the
magnitude of the velocities are present, since there is a residual
memory of the cold clumpy state. The precise form of the density profile
as well as the strength of the velocity correlations depend on the
details of the initial random positions of the clump centers, i.e. on
the seed of the random number generator.

\begin{figure}
  \plottwo{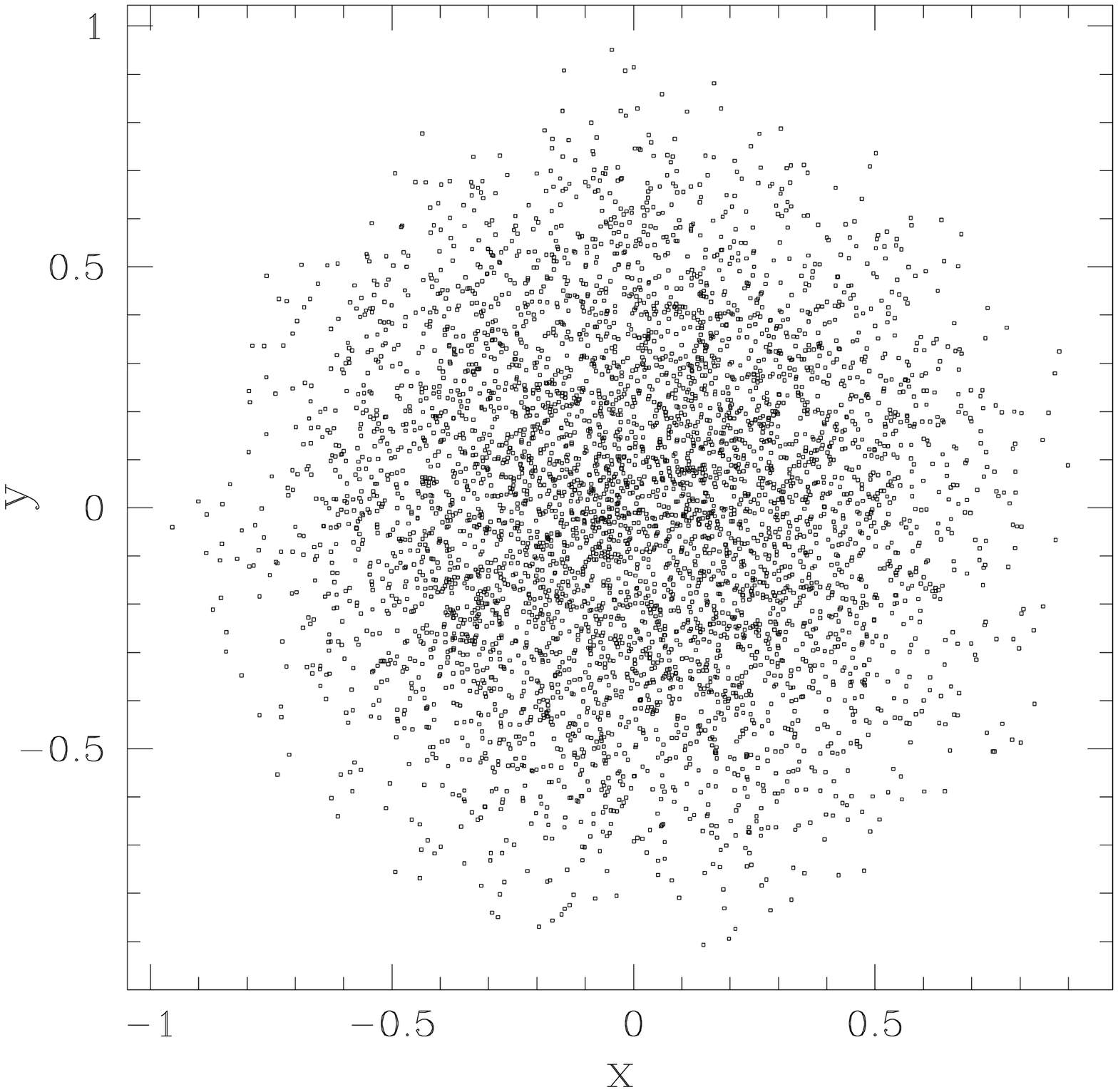}{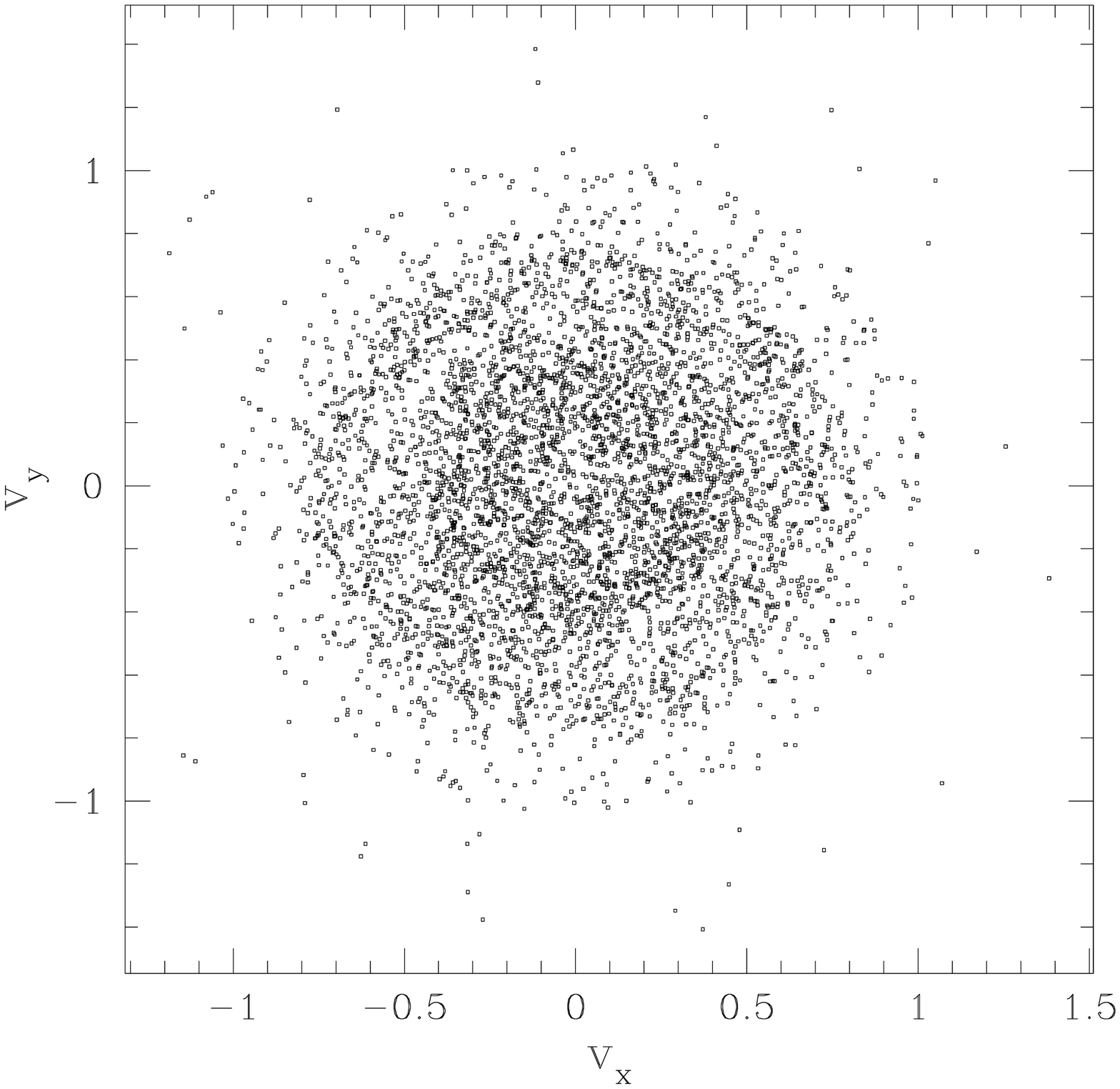}
  \caption{Snapshot of the initial conditions in the planes $(x,y)$
   (left panel) and $(v_x,v_y)$ (right panel) for simulation
   $S1$. The plots are given in code units. The effects of the
   symmetrization over the initial conditions of the kind considered
   in TBvA05 are easily appreciated in velocity space, where
   clumpiness before symmetrization was high (see Fig. 3 in TBvA05);
   here particles tend to form ring structures. The adopted
   symmetrization of initial conditions suppresses mixing in phase
   space during collisionless collapse.}\label{fig:IC}
\end{figure}

%%%%%%%% 
A snapshot of typical initial conditions from which our set of
simulations of collisionless collapse starts is reported in
Fig.~\ref{fig:IC}. The correlations in the magnitude of the velocities,
characteristic of the original clumps from which the present initial
conditions are extracted, can be seen as ring patterns created by the
particles in the $(v_x;v_y)$ space. We should emphasize that these
initial conditions are rather artificial and have been employed with
the specific purpose of studying the effects of the suppression of
mixing in phase space. It appears unlikely that in a realistic
scenario of galaxy formation the initial state is characterized by
such high degree of symmetry. Nevertheless, the properties of the
end-products of these collapse simulations are not too unrealistic, at
least in their outer parts (see Subsection 2.1).
%%%%%%%%

In Table~\ref{tab:S} we summarize the properties of this set of
simulations. The initial conditions of the $S1$ series are
generated using the same positions for the clump centers but a
different total number of particles $N$ (to ensure that the
properties of the end products do not depend on $N$). Run $S1^+$
has $8 \times 10^5$ simulation particles; $S1^*$ and
$S1^{\dagger}$ have $10^5$ particles, but the first simulation is
run with our particle-mesh code, while the second with Dehnen's
code. Runs $Sa$ and $Sb$ are colder versions  of $S1^*$ (obtained
by means of a global rescaling of the initial velocities by a
constant factor). Runs $S2$ and $S3$ are generated using a
different seed for the random numbers. In order to characterize
the level of anisotropy achieved in the end-states obtained from
the simulations, we refer to the global anisotropy parameter
$2K_r/K_T$ and to the anisotropy profile $\alpha(r)$, defined as
$\alpha(r) = 2 - (\langle v^2_{\theta}\rangle + \langle
v^2_{\phi}\rangle)/\langle v^2_r\rangle$.

\begin{figure}
  \plottwo{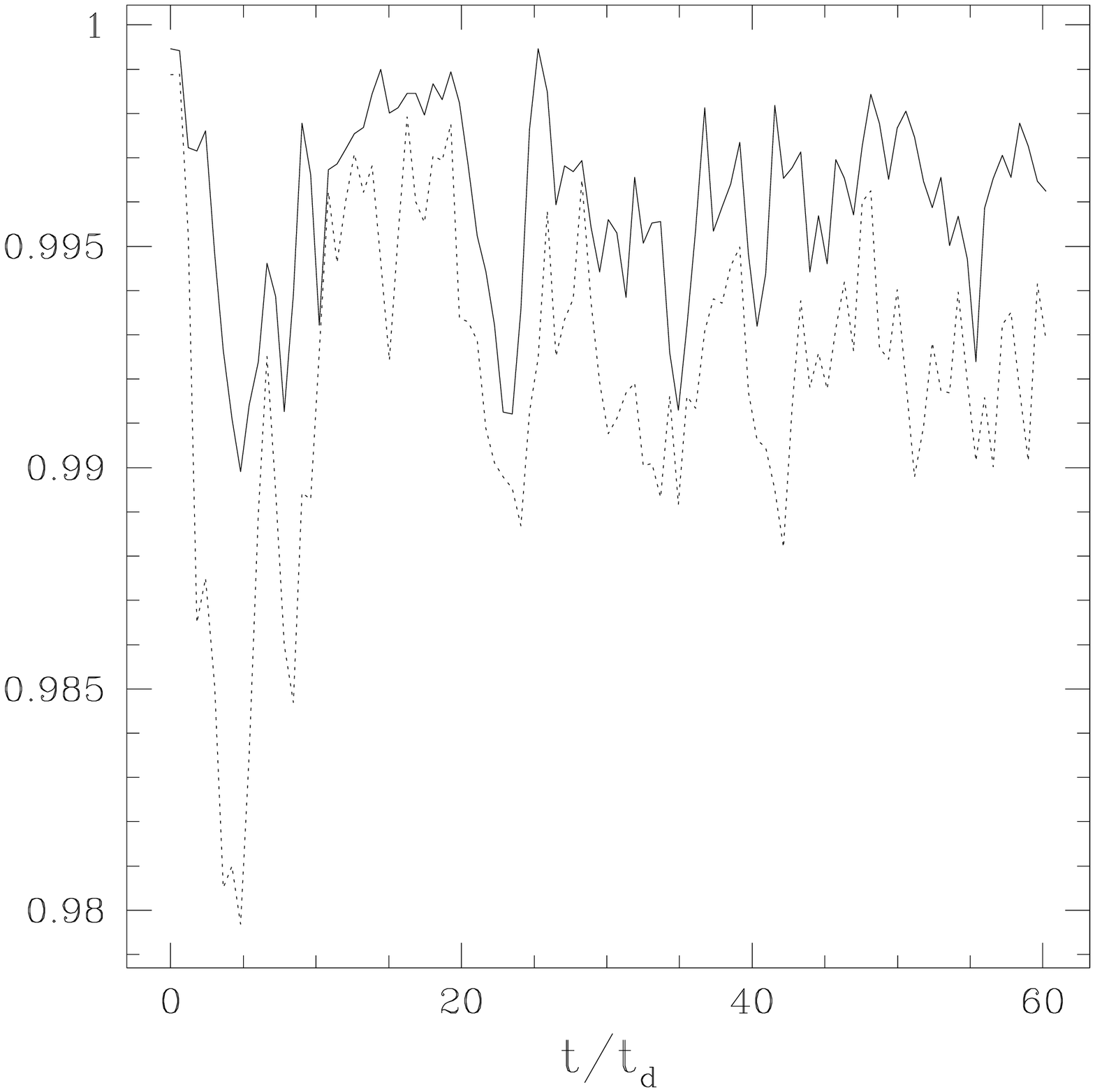}{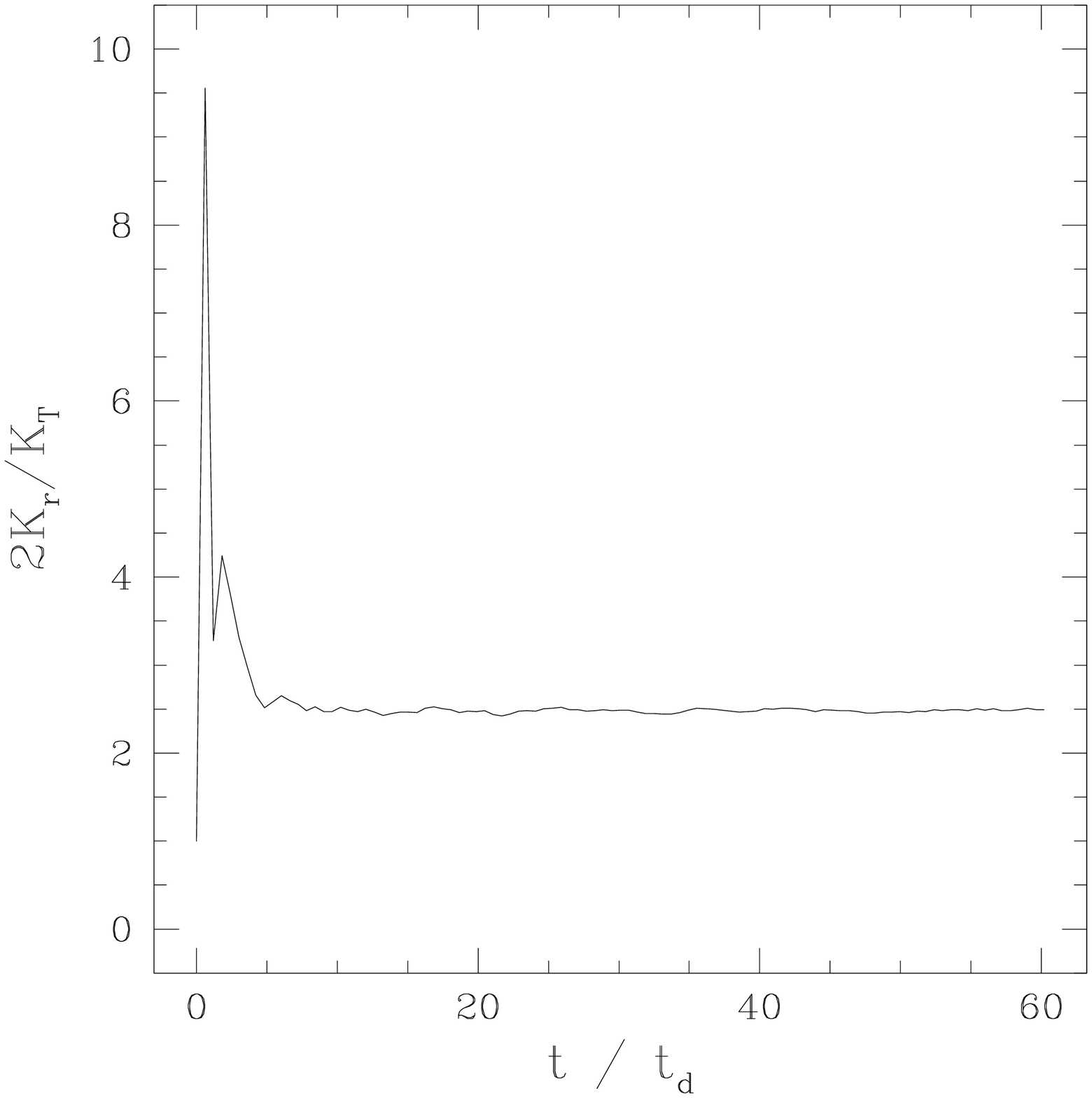}
  \caption{Left panel: evolution of the ellipticities $\epsilon=b/a$
  and $\eta=c/a$, where $a \geq b \geq c$ are the values of the axes
  of the system evaluated from the inertia tensor, for the $S1^+$
  simulation. Right panel: evolution of the ratio $2K_r/K_T$ for the
  same simulation. The dynamical time is defined in terms of the total
  mass and of the total energy as
  $t_d=GM^{5/2}/(-2E_{tot})^{3/2}$.}\label{fig:ellip}
\end{figure}

\begin{figure}
  \plottwo{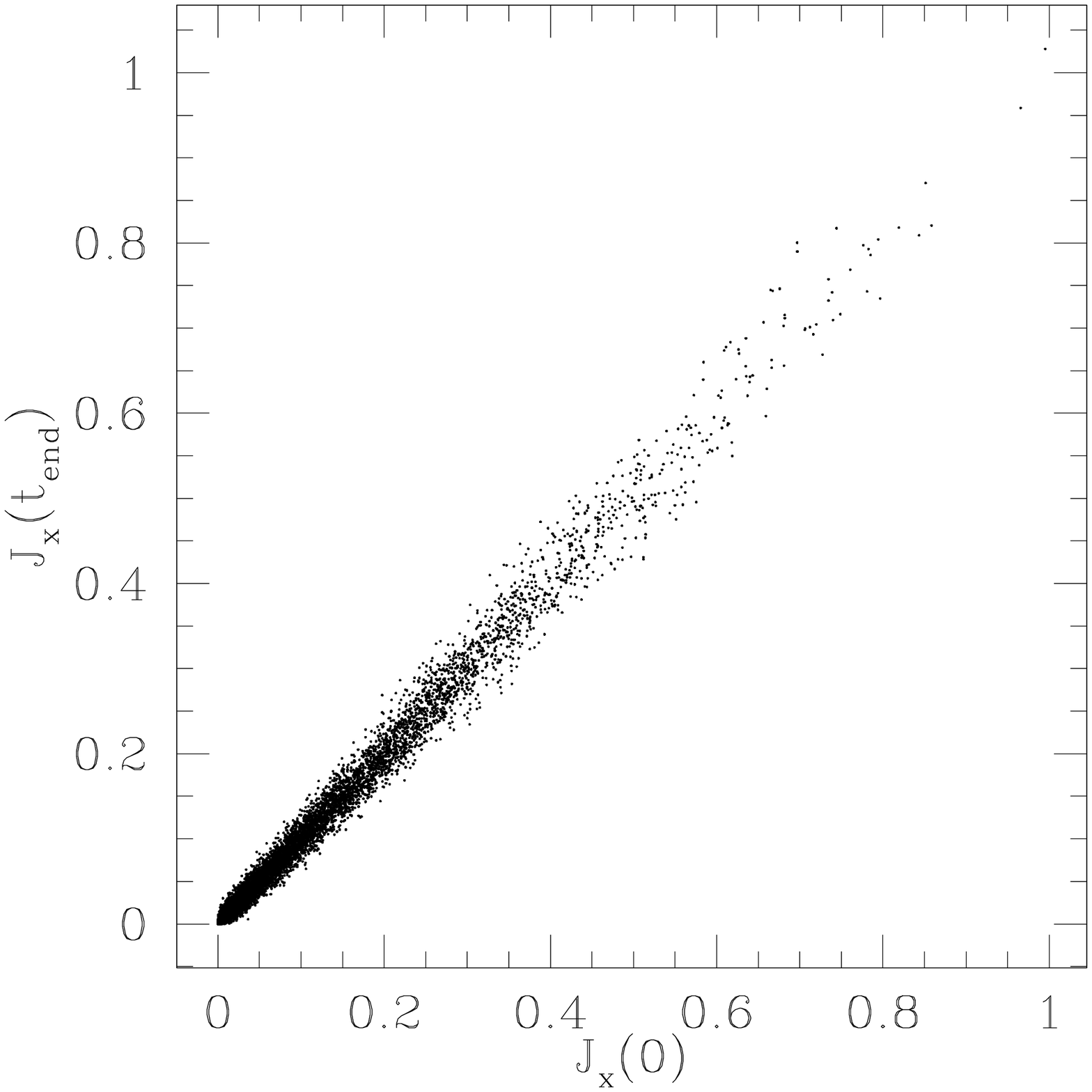}{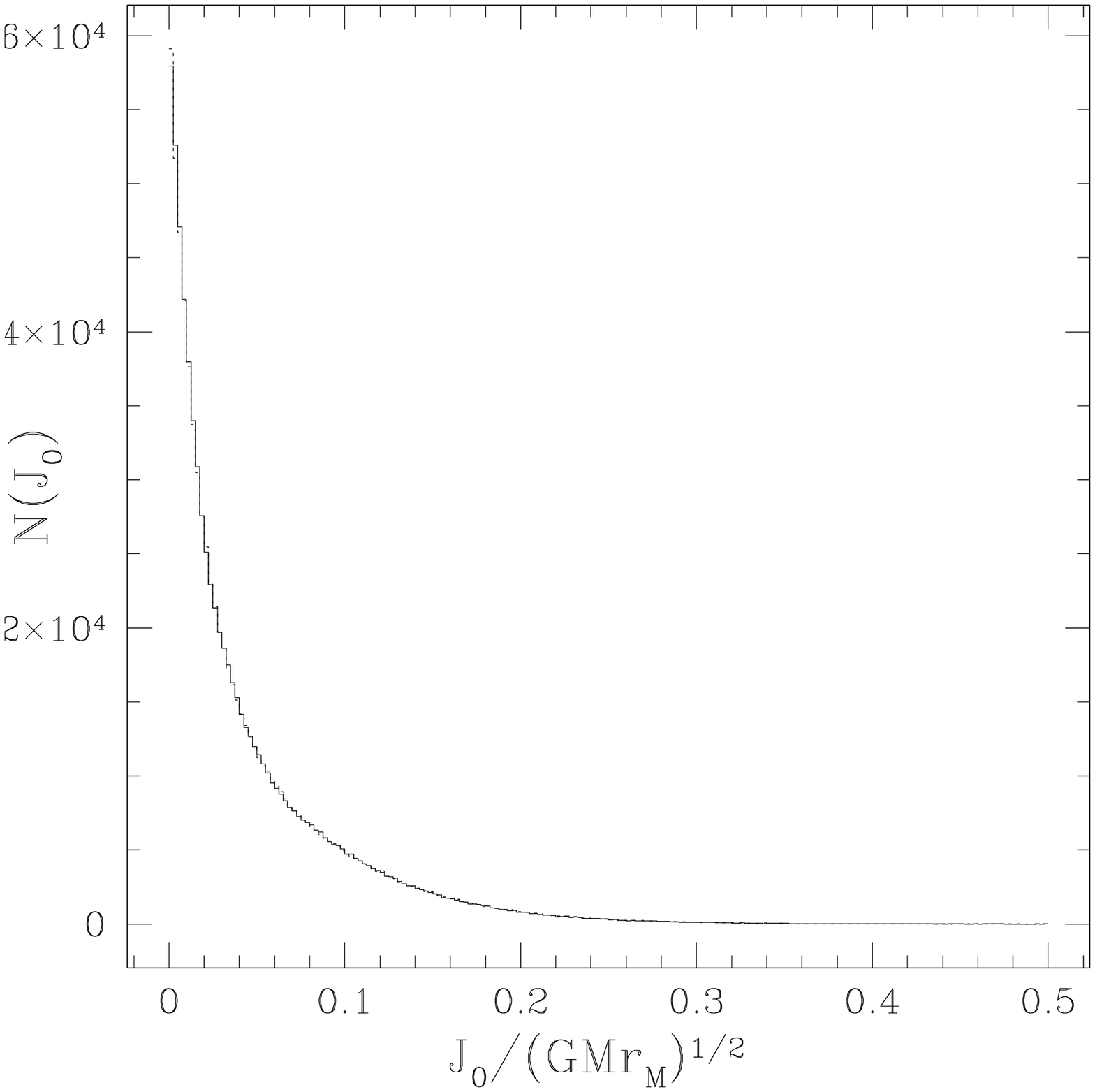}
   \caption{
%  Left panel: variation of one component of the specific
%  angular momentum of the simulation particles ($\Delta J_x =
%  J_x(t_{end}) - J_x(t=0)$) relative to a reference scale, defined as
%  the average value of $|J_x|$, for run $S1^+$. The single
%  particle energy $E$ is given in code units.
  Left panel: variation of one component of the specific angular
  momentum of the simulation particles for run $S1^+$.  Since during
  evolution the system remains spherically symmetric, the single
  particle angular momenta are approximately conserved. Right panel:
  Distribution of initial angular momenta (solid line) for the same
  simulation: the distribution basically coincides with that for the
  final angular momenta (dotted line). }\label{fig:Jsca}
\end{figure}

\begin{figure}
  \plottwo{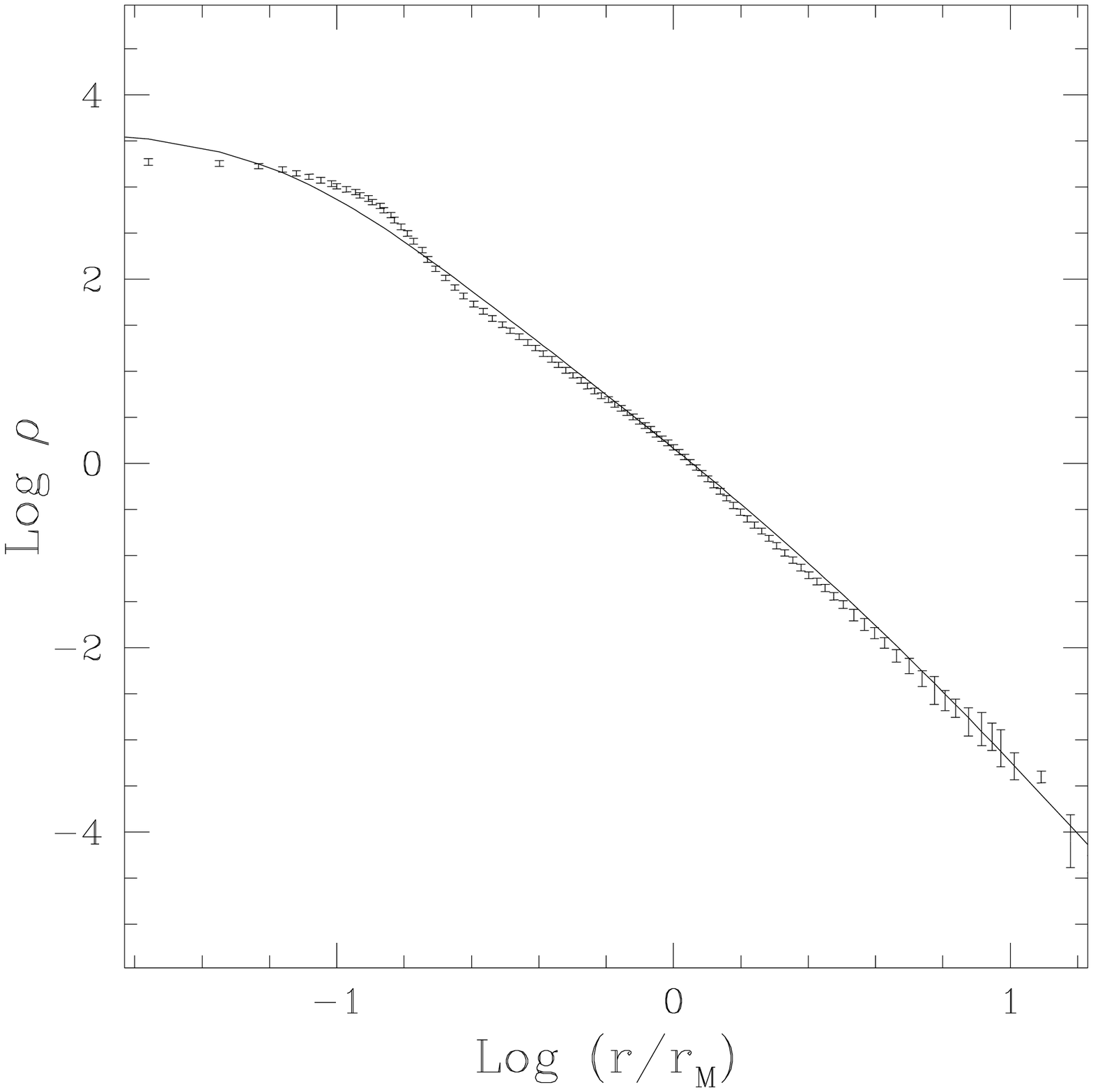}{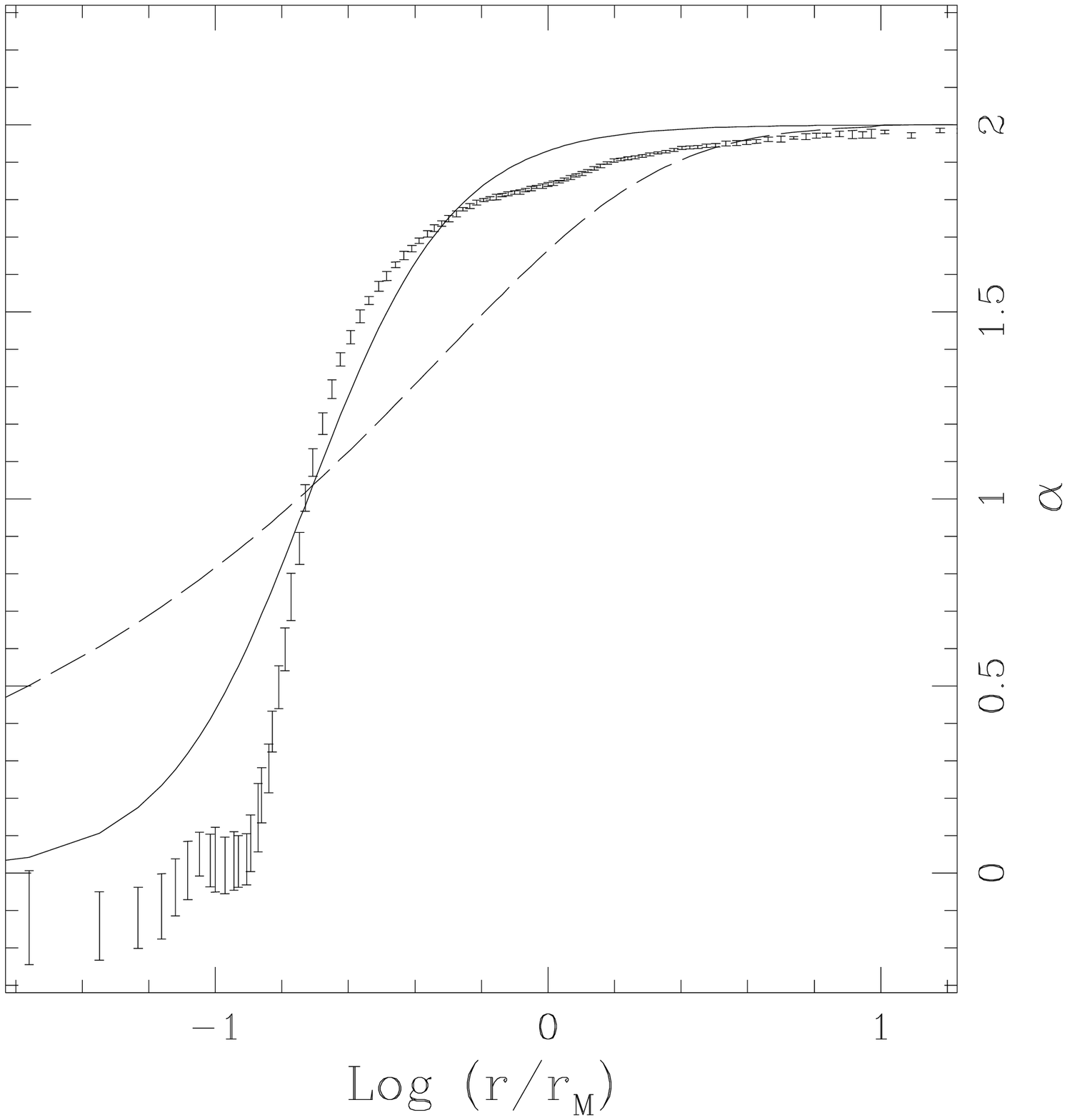}
  \caption{Density (in code units; left) and anisotropy profile
  (right) for the final state reached in the $S1^+$
  simulation. The solid line in the left panel is the density profile
  associated with the $(1/2;6.4)$ $\f$ model (see TB05 for definitions
  and notation). The anisotropy profile is compared to the
  Osipkov-Merritt anisotropy profile (solid line) and to the profile
  for an $\f$ model characterized by a similar amount of global
  anisotropy (dashed line; the $(1/2;1)$ model, which has the same
  $r_{\alpha}$ as the end-state for $S1^+$). Clearly the theoretical
  models are unable to capture the rapid increase in $\alpha$ at $r
  \sim r_{\alpha}$.}\label{fig:prof}
\end{figure}

During collapse, the spherical symmetry is well preserved, as can
be seen for $S1$ not only from the evolution of the eigenvalues of
the inertia tensor of the system (Fig.~\ref{fig:ellip}), but also
from the conservation of the single particle angular momenta
(Fig.~\ref{fig:Jsca}). Mass loss (i.e. the number of particles
that acquire a positive energy during the collapse) is limited,
well below the loss recorded for homogeneous uniform spheres with
similar initial virial ratio $u$, where the system can lose up to
one third of its total mass. The combination of spherical symmetry
and limited mass loss leads to high final density concentrations,
with $\rho(0)/\rho(r_M) \gtrsim 1500$ in run $S1$. As shown in
Fig.~\ref{fig:prof}, the density profile is reasonably well
represented by a rather concentrated $\f$ model (see TB05) or by a
Jaffe density profile \citep{jaf83}.

The global amount of pressure anisotropy (see Fig.~\ref{fig:ellip})
evolves rapidly in the first few dynamical times and then reaches
its quasi-equilibrium value at $t \approx 5 t_d$. When $2K_r/K_T$
reaches its peak value, at the time of maximum contraction of the
system, the anisotropy radius $r_{\alpha}$ (defined implicitly by
$\alpha(r_\alpha) = 1$) is located well inside:  at that time, for
run $S1$ the mass within $r_{\alpha}$ is only $1\%$ of the total
mass, while later the sphere associated with the anisotropy radius
contains approximately $20 \%$ of the total mass. The final global
content of pressure anisotropy is high also for run $S3$, which
starts from moderately warm initial conditions ($u=0.25$; for
comparison, see the results of non-symmetrized runs with
phase-space mixing presented in TBvA05).

Interestingly, the central regions have a final pressure anisotropy
profile slightly biased toward tangential orbits (see
Fig.~\ref{fig:prof}). This effect appears in the high resolution
$S1^+$ simulation, with $8 \times 10^5$ particles; the realization
with $10^5$ particles ($S1^*$) does not exhibit this feature and
indeed is characterized by a slightly higher value of $2K_r/K_T$. In
any case the transition from isotropic to radial pressure is very
sharp. The shape of the anisotropy profile cannot be represented
either by a profile similar to those of the $\f$ models or by an
Osikpov-Merritt profile \citep{osi79,mer85}.

To check the robustness of the numerical results, with Dehnen's tree
code we have let the final configuration reached in $S1^*$ evolve for
$30$ additional dynamical times, without noticing any sign of
significant changes. In addition, we have run a collapse simulation
starting with the same initial conditions as $S1^*$ using Dehnen's
tree code during the entire simulation (run $S1^{\dagger}$); this
simulation shows no macroscopic differences with respect to $S1^*$
(these are summarized in Table~\ref{tab:S} and are at the level of
$1\%$ ).

If we consider colder and colder initial conditions within the
framework of simulations considered in this paper, the radial
orbit instability eventually sets in. A reduction of the initial
virial ratio $u$ below $0.05$ for $S1$-like initial conditions (by
rescaling the velocities by a constant factor) leads to runs that
show evidence for the radial orbit instability: a simulation with
$u = 0.03$ ($Sb$) leads to a strongly flattened system, with a
final amount of global anisotropy close to $2.5$.

\begin{figure}
  \plottwo{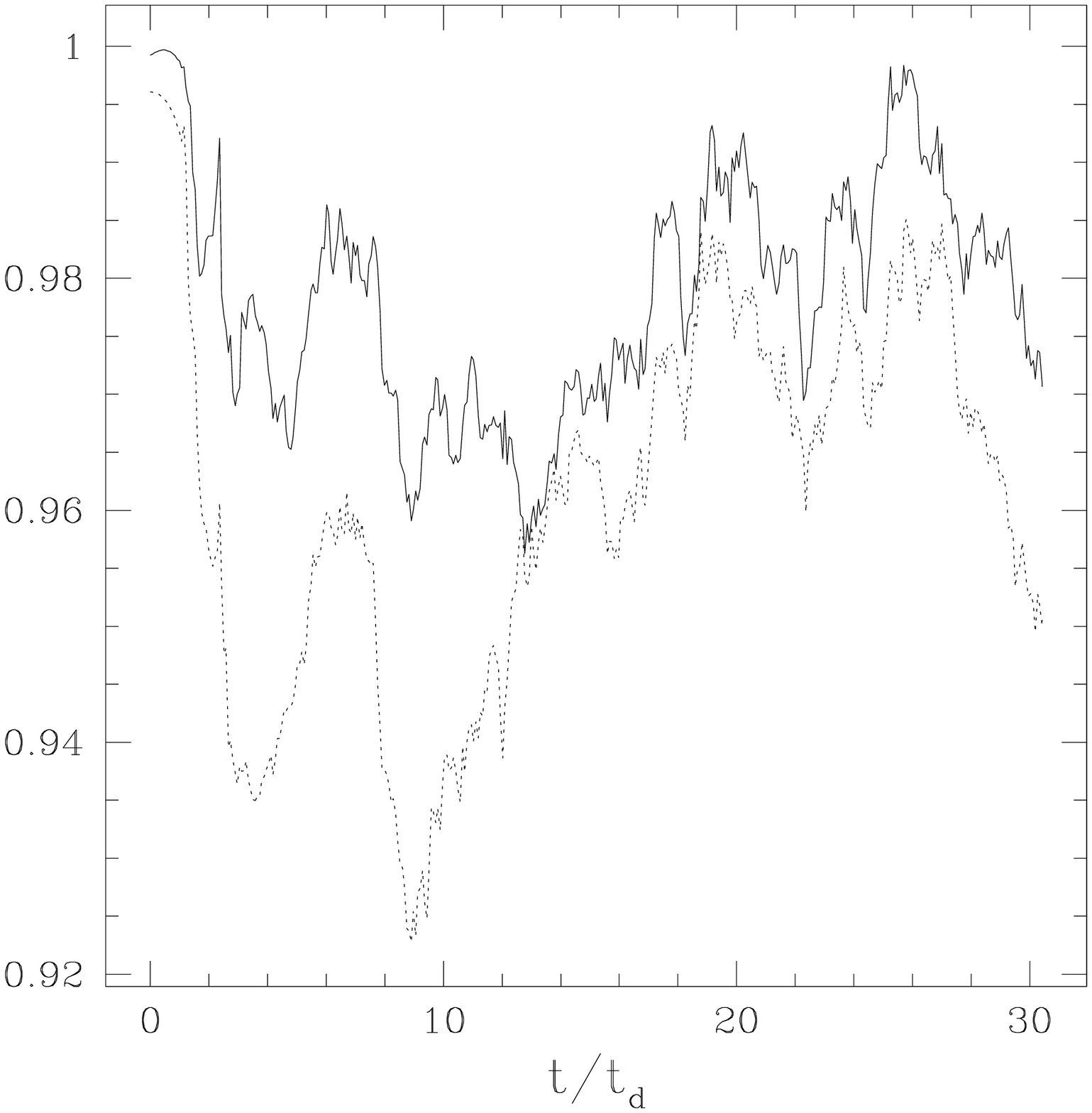}{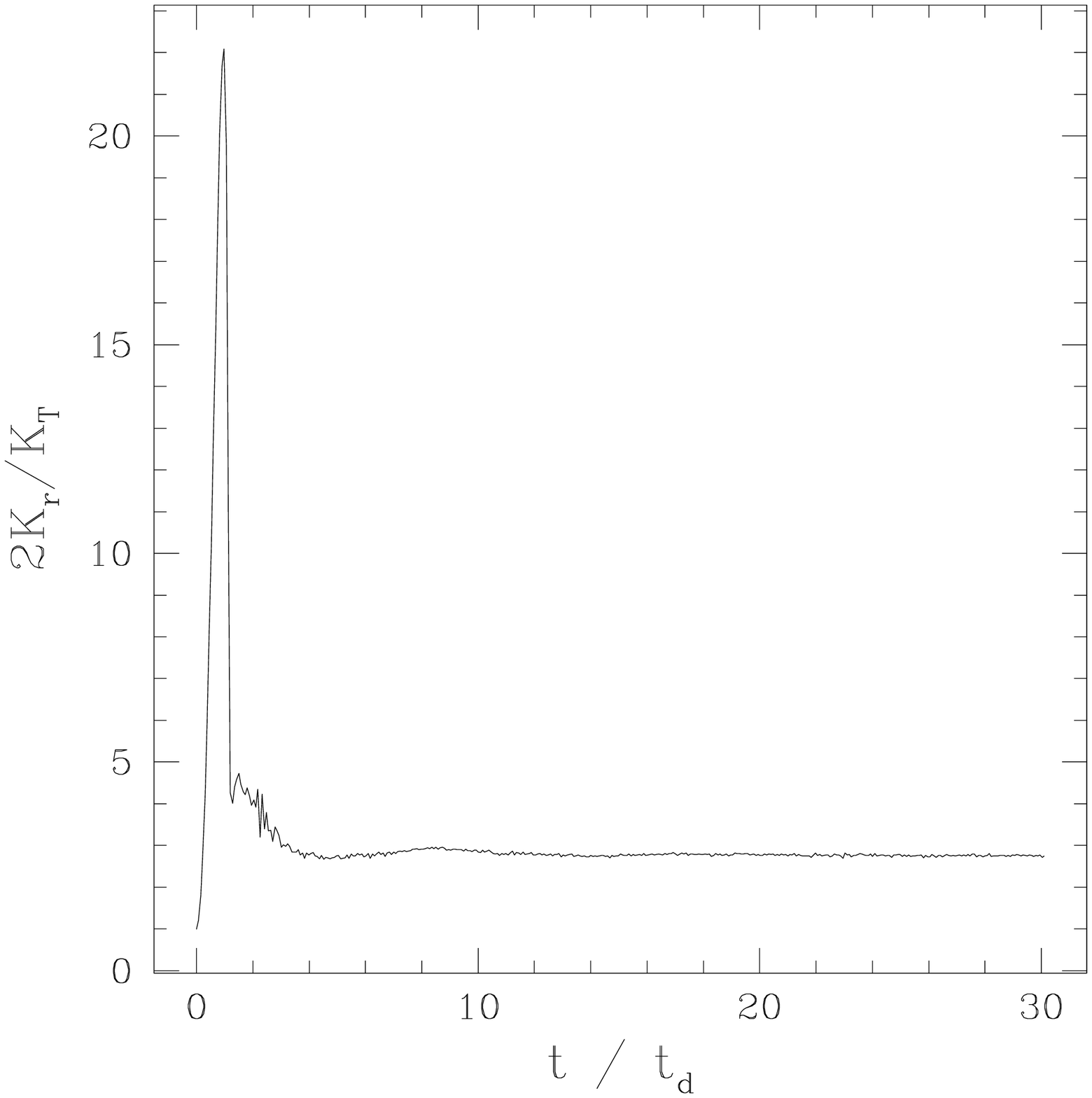}
  \caption{Ellipticities (left) and anisotropy content (right)
  for the $Sa$ simulation, as plotted in Fig.~\ref{fig:ellip}. Note
  the extremely high value of anisotropy content ($2K_r/K_T > 20$)
  reached during the collapse.}\label{fig:oscill}
\end{figure}

A simulation starting from $u = 0.05$ shows an interesting
behavior which we may identify as that of marginal stability,
since the final state is characterized by an aspect ratio (see
Fig.~\ref{fig:oscill}) $\eta$ that oscillates between 0.99 and
0.93, on a time scale longer than the dynamical time; the related
anisotropy content is high ($2K_r/K_T \approx 2.75$).

%%%%%%%% 
\subsection{Elliptical galaxies and the properties of the end-products
of some symmetric collisionless collapse simulations}

\begin{figure}
  \plotone{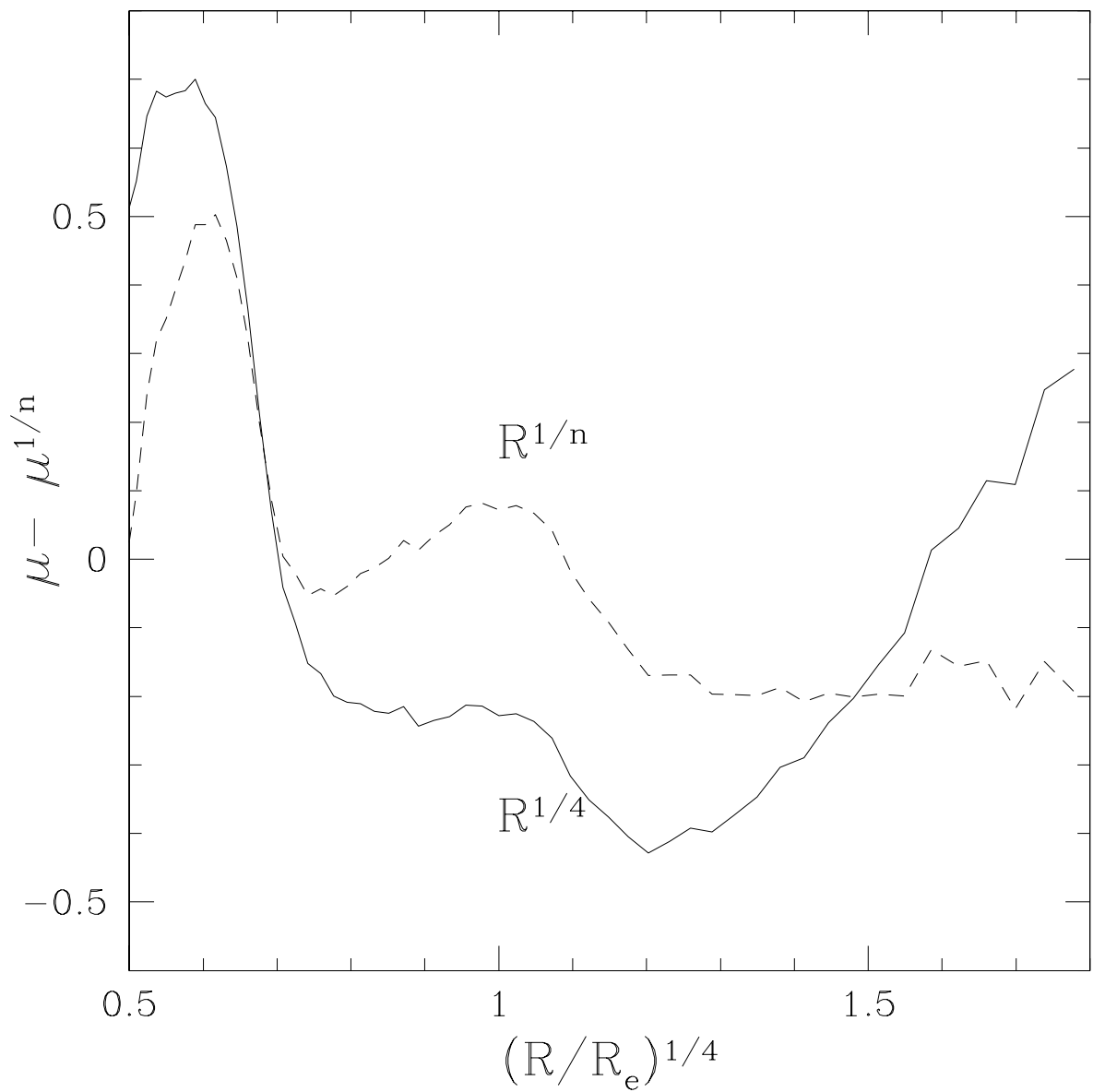}
   \caption{Residuals from the fit with the $R^{1/4}$ law (solid line)
   and with the $R^{1/n}$ law (dashed line), in magnitudes, of the
   projected density profile for the final configuration reached in
   simulation $S1^+$. The fit has been obtained by considering a constant
   mass-to-light ratio, at fixed effective radius $R_e$. The best
   fitting index $n$ is $n \approx 8.3$. Note the sizable residuals 
   in the central regions.}\label{fig:R14}
\end{figure}

As we have seen above, the end-products of this set of simulations,
starting from highly symmetric initial conditions, are characterized
by a density profile which is not too far from that of models, such as
the \citet{jaf83} profile, that are known to be associated with
realistic surface brightness profiles for bright elliptical galaxies
(under the assumption of constant mass-to-light ratio). To quantify
the effects of the systematic differences between these models and the
products of the simulations, in Fig.~\ref{fig:R14} we consider the
projected density profile measured at the end of simulation $S1^+$ and
we compare it to the $R^{1/4}$ \citep{dev48} and to the $R^{1/n}$
\citep{ser68} laws. The residuals from the best fit, at fixed
effective radius $R_e$, are not too large, especially for the Sersic
law with $n \approx 8.3$. In fact, the deviations are within $0.2$
magnitudes over a wide interval in radius, from $0.2$ to $10~R_e$;
however, in the central regions the difference can be as high as $0.5$
magnitudes, probably because of the presence of the ``bump" in the
density profile around $0.1~r_M$ (see Fig.~\ref{fig:prof}). Residuals
from the $R^{1/4}$ are larger. 
%This may be taken as a confirmation
%that artificial initial conditions may lead to unrealistic
%end-products. ## mt: non mi piace questa frase... alla fine i prodotti
%non sono cosi' irrealistici. 

For a direct comparison of the models considered in this paper and the
observations, one might think of making a kinematical
test. Unfortunately, as to the anisotropy of the velocity dispersion
tensor, at present a comparison with the observations would be extremely
difficult to carry out.  The best empirical evidence for radial
anisotropy in elliptical galaxies comes from the observed flattening
(and triaxiality) in objects that are not rotationally supported. In
principle, pressure anisotropy affects the shape of the velocity
distribution integrated along the line of sight, so that its presence
could be detected by observing the shapes of the line profiles (and in
particular the deviations from Gaussian profiles). In practice, the
effects on line profiles are rather modest even for significant amounts
of pressure anisotropy, so that, within the uncertainty limits of the
measurement, the pressure anisotropy profile of elliptical galaxies
remains basically undetermined by current observations \citep[e.g.,
see][]{ger93,ger01,deZee02}.  In other words, it would be very hard to
ascertain whether some ellipticals are indeed as isotropic in their
inner regions as implied by some of the models that we have
investigated.

As briefly mentioned in the Introduction, numerical simulations of
galaxy formation suggest that elliptical galaxies may be close to
marginal stability with respect to the radial orbit instability, close
to the threshold stated by Polyachenko \& Shukhman (1981). In this
respect, studying the effects on the radial orbit stability of a
highly isotropic core, as we do in this paper, is of interest, because
there are astrophysical processes that are often ignored in the
simulations of galaxy formation and should operate in real elliptical
galaxies, which may favor a more isotropic core or even an anisotropic
core with a slight bias of the pressure tensor in the tangential
directions. Here we have in mind processes such as those due to the
presence of a super-massive black hole (for the effects of a black
hole on the pressure anisotropy within the sphere of influence see
e.g., \citealt{cip94}, \citealt{bau04}). To be sure, for
a physically significant application of these ideas, one should check
the time-scales of the competing processes that are invoked, before
drawing the relevant conclusions.

\section{Exceptionally stable equilibria constructed from the Jeans equations}\label{sec:rho_alpha}

We now address the issue of whether a given initial configuration,
characterized by assigned density $\rho(r)$ and pressure
anisotropy $\alpha(r)$ profiles, is stable with respect to the
radial orbit instability. We are not aware of models with
analytical distribution function able to incorporate the sharp
feature in the anisotropy profile, of the kind observed at the end
of the simulations described in the previous Section. Therefore,
we decided to initialize the simulations by means of candidate
equilibrium solutions obtained from the Jeans equations, as
outlined in the Appendix. We should emphasize that the simulations
that we describe below in this Section are simulations of {\it
candidate} equilibria. In fact, since we never reach the point of
actually reconstructing an underlying equilibrium distribution
function, if we happened to find significant evolution we could be
either in a situation of genuine instability, or, more simply, in
a situation of non-equilibrium. In turn, since we will show cases
where we do {\it not} find such evolution, we may claim that
indeed we have found not only a genuine quasi-equilibrium state
but also proved that it is approximately stable.

To model a density profile of the kind found at the end of the $S1$
simulations, when no evolution is observed over several dynamical
times, we use a superposition of a regularized \citet{jaf83}
profile:
\be\label{eq:jaff}
\rho_J(r) = \frac{A}{(r^2+\epsilon^2)(r+b)^2},
\ee

\noindent with $A$, $\epsilon$, and $b$ free scales, and a central
core of the form:

\be
\rho_C(r) = \frac{A^{\prime}}{(r^2+ \xi b^2)^{20}},
\ee
%

% VALUES FROM THE BEST FIT: 
%   A       = 0.036
%   epsilon = 0.022
%   b       = 0.27
%   xi      = 0.60
%   A'      = 1300 * (xi*b^2)**20

\noindent where $A^{\prime}$ is a free scale, while $\xi$ is a
dimensionless parameter of order $1$ ($\xi \approx 0.6$). The form for
the density $\rho_{mod} = \rho_J+\rho_C$ is taken for convenience, so
as to reproduce not only the large-scale structure of the density
distribution realized in the $S1$ simulations, but also the bump in
the density profile around $r_M/5$ (see Fig.~\ref{fig:prof}).

As noted earlier, the $S1$ anisotropy profile is flatter than the
corresponding Osipkov-Merritt profile with same $r_{\alpha}$ at
small radii, while it is less steep at large radii. Thus we chose
to represent the profile with:

\be
\alpha(r) = 2 \frac{r^{\gamma}}{r^{\gamma}+r_a^{\gamma}}
\ee
\noindent with $r_a$ and $\gamma$ being free parameters. A single
choice of the $(r_a;\gamma)$ values is unable to correctly reproduce
the anisotropy profile measured from the simulation over the entire
radial range. Thus we fit separately the anisotropy in the ``core'',
up to a radius $r_{ch} \gtrsim r_{\alpha}$, where $\gamma = 4$, and in
the ``halo'', i.e. for $r > r_{ch}$, where $\gamma = 6/5$. In a
neighborhood around $r_{ch}$ the two profiles are matched so that the
final $\alpha(r)$ and its derivative are continuous functions.

With a suitable choice for the various parameters that define the
above functions, the profiles obtained from the simulation can be
fitted with an accuracy of better than $10$ percent. We take this
as a good starting point to investigate the stability of
equilibrium configurations similar to those produced in the
simulations of collisionless collapse.

\begin{figure}
  \plottwo{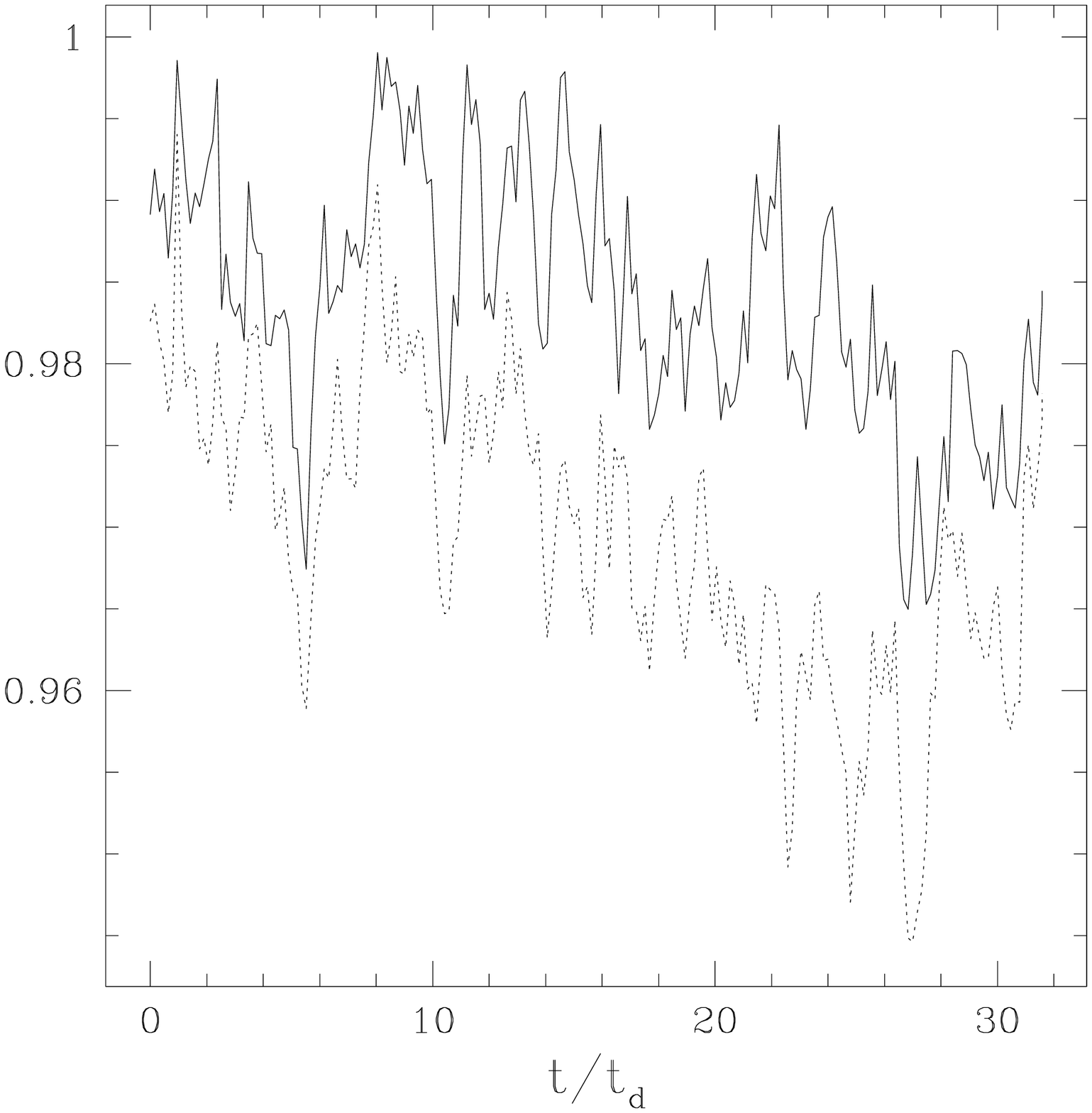}{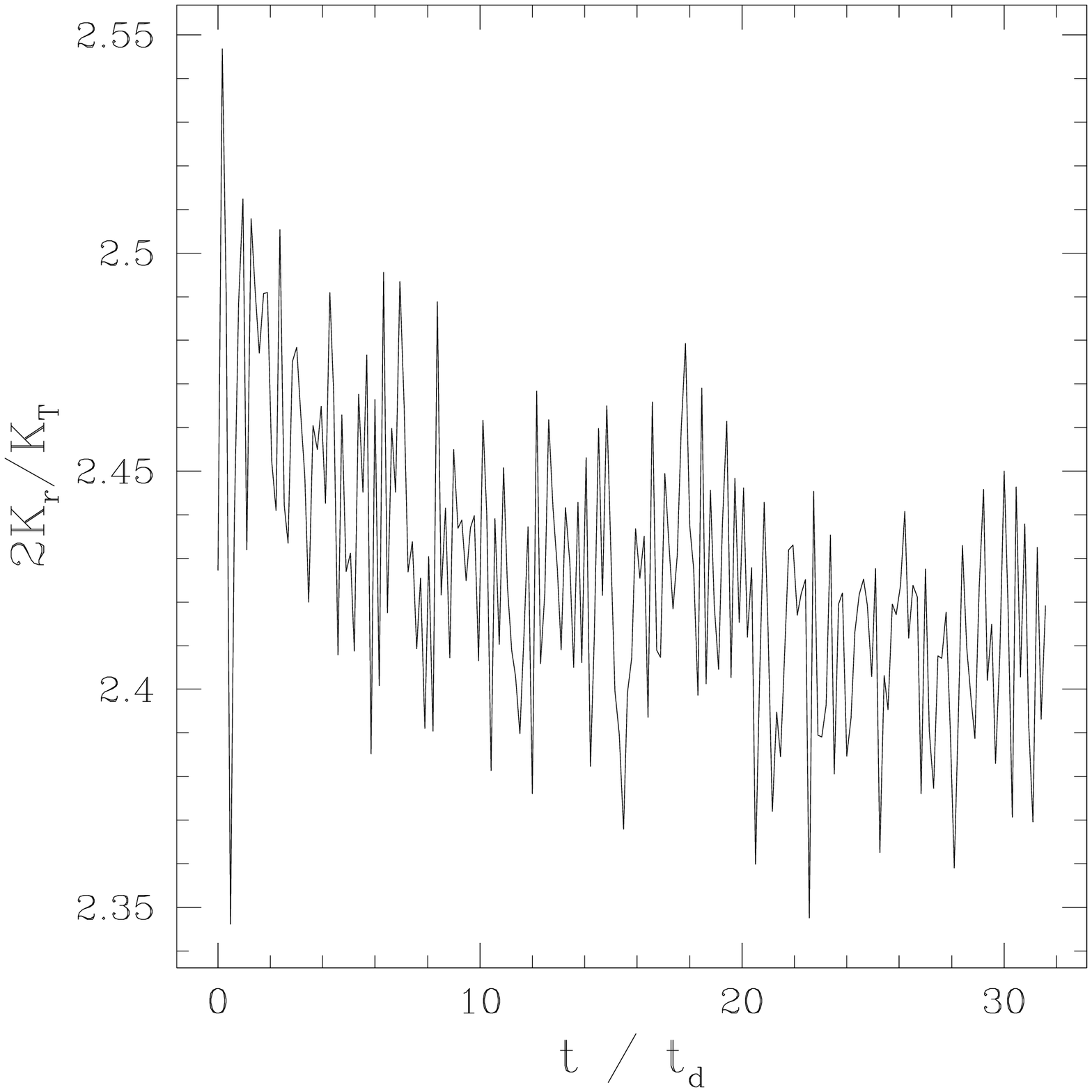}
  \caption{Ellipticities (left) and anisotropy content (right)
  for the $J1$ simulation, as plotted in Fig.~\ref{fig:ellip}.}
  \label{fig:ani_j1}
\end{figure}

We have first studied the evolution of a model initialized with
density and anisotropy profiles similar to those of runs $S1$ (see
entry $J1$ in Table~\ref{tab:jeans} and Fig.~\ref{fig:ani_j1}). We
have then proceeded to study the evolution of neighboring
configurations by slightly modifying the density and/or the anisotropy
profiles. In particular, we have considered models with a density
profile without the inner ``bump" (i.e., without the $\rho_C$
contribution), and different forms of anisotropy profile, ranging from
steep profiles over the entire radial range, to Osipkov-Merritt and
$\f$-like profiles. Interestingly we have found that although none of
the various combinations turns out to be violently unstable,
nevertheless, the only simulation where practically no sign of
evolution occurs is $J1$, the one associated with the profiles that
best fit those of $S1$.

\begin{figure}
  \plottwo{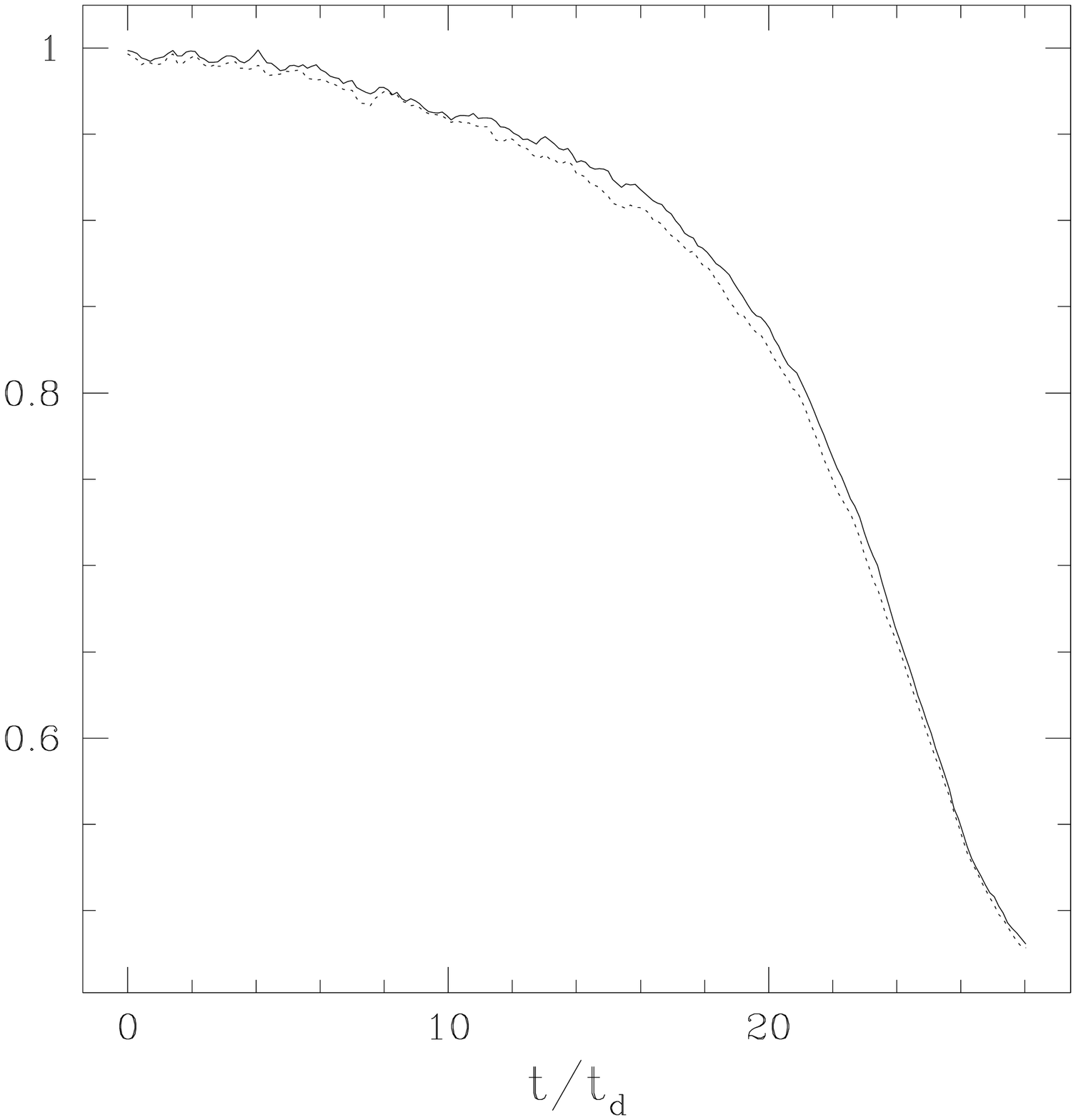}{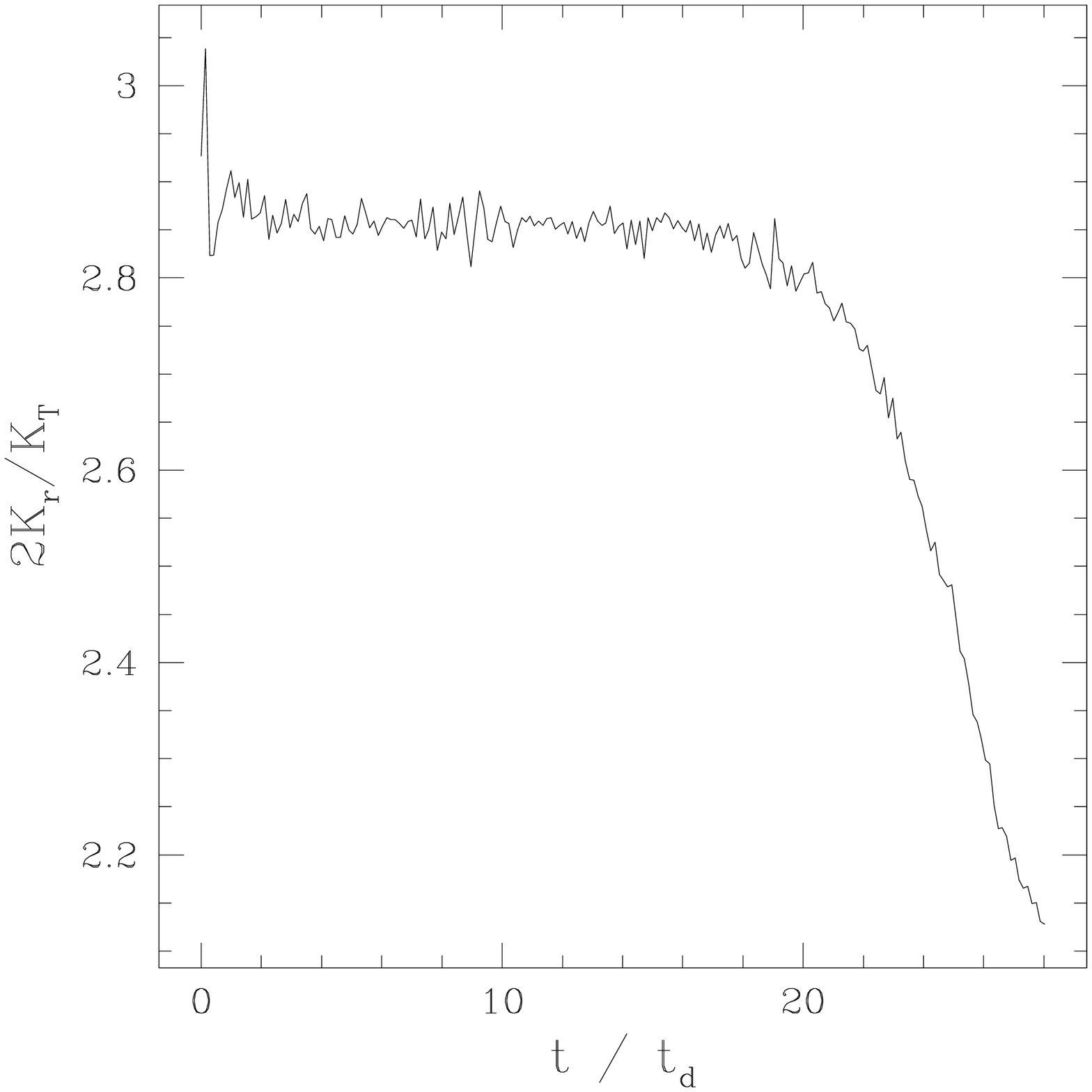}
  \caption{Ellipticities (left) and anisotropy content (right) for the
  $J2$ simulation ($2K_r/K_T = 2.92$ at $t=0$), as plotted in
  Fig.~\ref{fig:ellip}.}\label{fig:ellip_j2}
\end{figure}

For the systems that show definite signs of evolution, the
evolution appears to be very slow. For example, in the $J2$
simulation the initial configuration lasts basically unchanged for
more than $10$ dynamical times. In this case, illustrated in
Fig.~\ref{fig:ellip_j2}, the system remains
close to spherical symmetry with $2K_r/K_T \approx 2.9$, and ends
up only much later as a prolate system.

%%%%%%%%%%%%%%%%%%%%%%%%%%%%%%%%%%%%%%%%%%%%%%%%%%%%%%%%%%%%%%%%%%%%%%%%%%%%%%%%%
\section{Conclusions} \label{sec:con}

In this paper we have studied and clarified the dependence of the
radial orbit instability on the shape of the anisotropy profile in the
central region of a stellar system. By means of a series of numerical
simulations we have shown that stable, centrally isotropic equilibria
with a significant global amount of anisotropy can be reached during
highly symmetric cold collapse events or initialized by solving the Jeans
equations. In particular we have found a metastable state with
$2K_r/K_T \approx 2.9$.

\begin{figure}
\plottwo{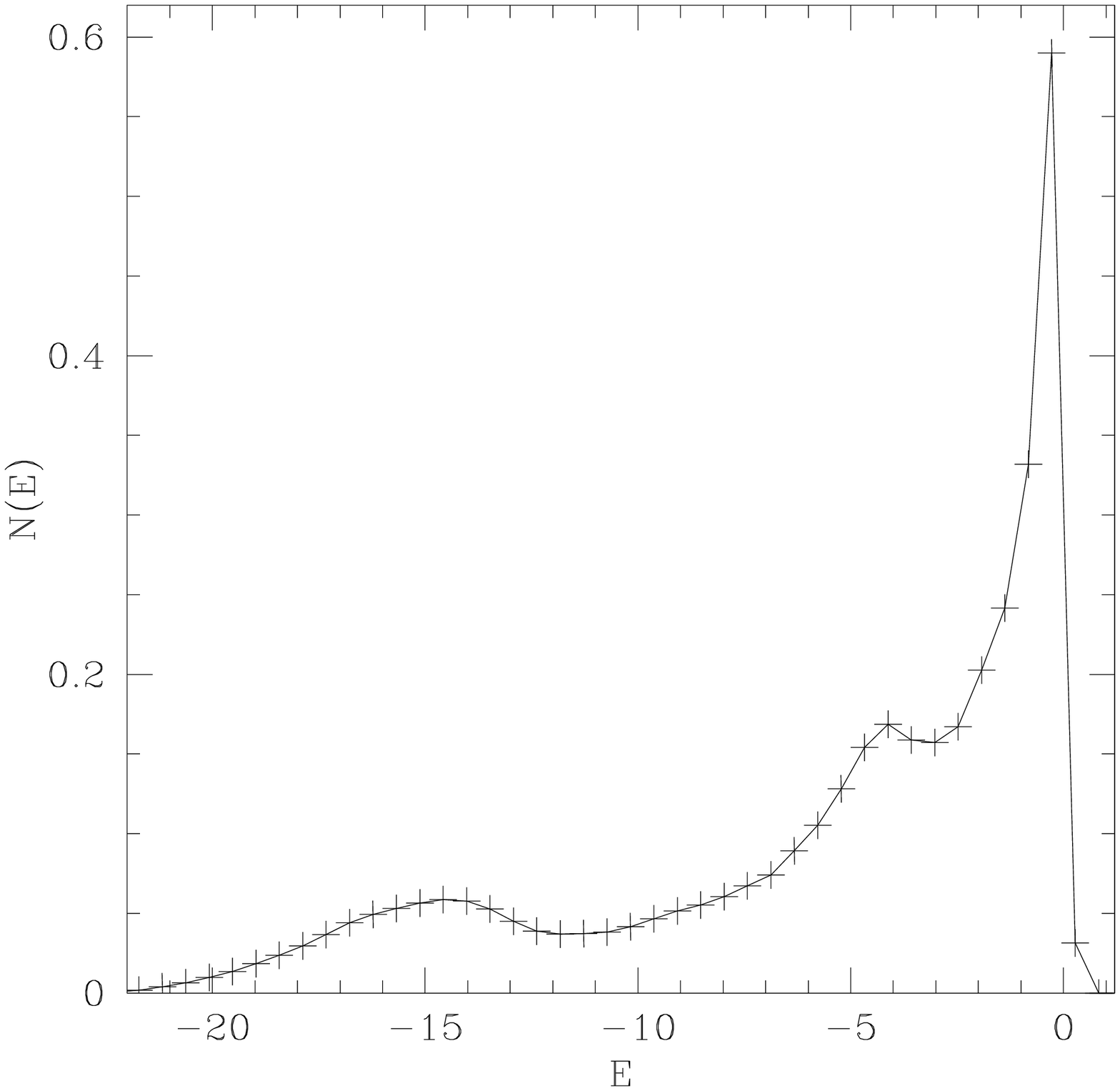}{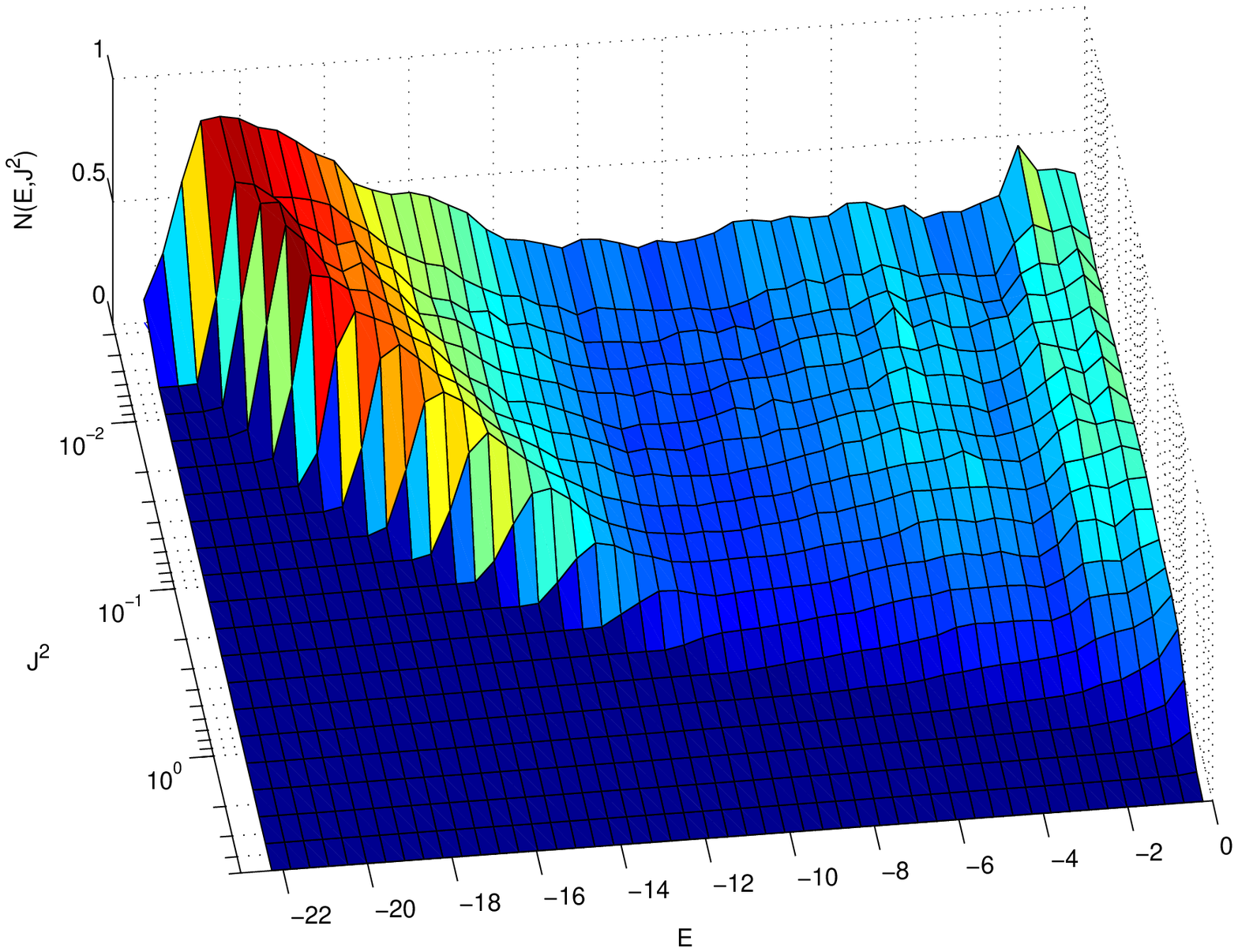}
\caption{Phase space distributions $N(E)$ (left) and
  $N(E,J^2)$ (right) for the final state reached in the $S1^+$
  simulation. The peak in the bottom left panel around
  $E \approx -14$ (in code units) is associated with the isotropic
  core.}\label{fig:profNEJ}
\end{figure}

The experiments that we have performed suggest that the presence of an
isotropic core may act as an important stabilizing factor for the radial
orbit instability. The two-component (core-halo) structure for run
$S1^+$ is indeed evident not only in the density and pressure anisotropy
profiles (see the structure out to $r \approx r_M/5$ in
Fig.~\ref{fig:prof}), but also in the phase space distribution $N(E)$,
as indicated by the peak located at high values of the binding energy
($E \approx -14$, in code units; see left panel of
Fig.~\ref{fig:profNEJ}).

One indication for the stabilizing effect of a highly isotropic core may
be found by recalling that the linear modal stability analyses have
shown that the density profiles of the fastest growing modes are peaked
in the innermost region of the system \citep[see e.g.,][ Fig
8]{ber94}. Thus we may argue that, if the core were almost perfectly
isotropic, there would be no room for the modes to be excited.

These results are qualitatively reminiscent (although the physical
mechanisms that are involved are completely different), of the
stabilizing role that a hot bulge can provide in relation to the
stability of self-gravitating disks, as noted in a number of papers
starting with \citet{ber77} and \citet{sel81}. For disks, the detailed
mechanisms underlying the origin or the suppression of global bar and
spiral instabilities are reasonably well understood and known to depend
on the key structural properties of the basic state (i.e., the disk
density, effective velocity dispersion, and differential rotation
profiles). For anisotropic spherical stellar systems we still lack a
clear picture of the relevant underlying mechanisms. In this respect,
this paper offers an interesting clue to more systematic studies that
should be devoted to investigating the nature of the radial orbit
instability in two-component systems, also in view of the central
properties of the dark halo.

%%%%%%%%%%%%%%%%%
\acknowledgments 

We would like to thank T.S. van Albada for a number of conversations and
useful suggestions. This work has been partially supported by the
Italian Ministry MIUR.

\appendix
\section[]{Candidate collisionless equilibria generated by the Jeans
equations}\label{sec:jeans}

The goal is to construct stellar dynamical configurations, as
initial conditions for N-body simulations, corresponding to
systems with given density and pressure anisotropy profiles. As is
well known, in general this problem is not well-posed, because a
given pair $(\rho(r), \alpha(r))$ may even lack a physically
acceptable underlying equilibrium distribution function. We
proceed by applying the necessary constraints of the Jeans
equations and then initialize our simulations with a procedure
described below, well aware that we may actually miss the desired
equilibrium conditions (see comment at the beginning of Sect.~3).

The knowledge of the density profile immediately identifies the
self-consistent potential $\Phi(r)$. In addition, it defines the
integrated mass profile, from which the positions of the particles
can be correctly initialized. In the absence of an analytical
distribution function, to complete the initialization of the
N-body collisionless candidate equilibrium configuration we then
resort to the Jeans equations to extract the information about the
velocity dispersion profile that would be required by the
conditions of equilibrium. In practice, for a non-rotating
spherically symmetric system the relevant hydrostatic equilibrium
equation can be written as:

\be \label{eq:ode} \frac{d (\rho(r) \sigma^2_r(r))}{dr} + \frac{\rho(r)}{r}
\alpha(r) \sigma^2_r(r) = - \rho(r) \frac{d \Phi(r)}{dr}, \ee

\noindent where $\Phi$ represents the mean field gravitational
potential (which can be taken to be known for known $\rho(r)$) and
$\sigma^2_r = \langle v^2_r \rangle$ the velocity dispersion in
the radial direction. For assigned profiles  $(\rho(r),
\alpha(r))$, under the natural boundary condition given by

\be
\rho(\infty) \sigma^2_r(\infty) = 0,
\ee

\noindent the equation can be solved for $\sigma^2_r$:

\be \label{eq:sol} \sigma^2_r(r) = \frac{1}{\rho(r)}
\int_r^{+\infty} dr^{\prime} \frac{d \Phi(r^{\prime})}{d
r^{\prime}} \rho(r^{\prime}) \exp{\left \{ \int_r^{r^{\prime}}
\frac{\alpha(\tilde{r})}{\tilde{r}} d \tilde{r}\right \}}. \ee

\noindent Thus we have obtained the velocity dispersion profiles
consistent with the given density and anisotropy profiles.

At this point, we initialize our N-body code by assuming that the
velocity dispersion profiles obtained above correspond, at least
approximately, to a truncated Gaussian distribution in velocity
space (the truncation is imposed by the requirement that only
bound particles, i.e. particles with negative energy, belong to
the system). This assumption gives a reasonable description of
systems for which the velocity distribution is determined by
physically motivated distribution functions (e.g., for the stable $\f$
models); but, in general, there is no guarantee that N-body
systems initialized by this method be in approximate dynamical
equilibrium.

In our case, the limitations of this approach are not severe,
because we are interested in a \emph{stability} study. If we
happen to obtain a configuration with no significant signs of
evolution, we are certain \emph{a posteriori} that a
quasi-equilibrium state has indeed been produced. On the other
hand, even if the initialization gives a configuration not in
dynamical equilibrium, but a nearby equilibrium configuration is
approached on a very short time scale (typically on the order of
one dynamical time), then we may argue that we have also obtained
our goal of producing a quasi-equilibrium state with
characteristics close to those desired.

\subsection{Tests on known distribution functions}

We have tested our method on isotropic systems by initializing
Plummer and polytropic models and then on some anisotropic
configurations associated with the $\f$ family (see TB05), under
conditions of stability. By construction, we have a good match at
the level of density, velocity dispersion, and anisotropy
profiles. Differences in the fourth order moment of the velocity
distribution are typically at the level of a few percent. The
models initialized with the method outlined above appear to be
stable, except for some modest re-arrangements of the initial
anisotropy profile in the case of the $\f$ models. In the latter
case, there is a tendency for the anisotropy content in the
central region to decrease so that a flatter profile in
$\alpha(r)$ is eventually generated.

%%%%%%%%%%%%%%%%%%%%%%%%%%%%%%%%%

\clearpage

% Figures here:

%%%%%%%%%%%%%%%%%%%%%%%%%%%%%%%%%%%%%%%%%%%%%
%%%%%%%%%%%%%%%%%%%%%%%%%%%%%%%%%%%%%%%%%%%%%%%%

\clearpage

%%%%%%%%%%%%%%%%%%%%%%%%%%%%%%%%%%%%%%%%%%%%%
%%%%%%%%%%%%%%%%%%%%%%%%%%%%%%%%%%%%%%%%%%%%%%%%

\clearpage

%%%%%%%%%%%%%%%%%%%%%%%%%%%%%%%%%%%%%%%%%%%%%
%%%%%%%%%%%%%%%%%%%%%%%%%%%%%%%%%%%%%%%%%%%%%%%%

\clearpage

%%%%%%%%%%%%%%%%%%%%%%%%%%%%%%%%%%%%%%%%%%%%%

\clearpage

%%%%%%%%%%%%%%%%%%%%%%%%%%%%%%%%%%%%%%%%%%%%%
%%%%%%%%%%%%%%%%%%%%%%%%%%%%%%%%%%%%%%%%%%%%%%%%

\clearpage

%%%%%%%%%%%%%%%%%%%%%%%%%%%%%%%%%%%%%%%%%%%%%
%%%%%%%%%%%%%%%%%%%%%%%%%%%%%%%%%%%%%%%%%%%%%%%%

\clearpage

%%%%%%%%%%%%%%%%%%%%%%%%%%%%%%%%%%%%%%%%%%%%%
%%%%%%%%%%%%%%%%%%%%%%%%%%%%%%%%%%%%%%%%%%%%%%%%

\clearpage
%%%%%%%%%%%%%%%%%%%%%%%%%%%%%%%%%%%%%%%%%%%%%
%%%%%%%%%%%%%%%%%%%%%%%%%%%%%%%%%%%%%%%%%%%%%%%%

\clearpage

%%%%%%%%%%%%%%%%%%%%%%%%%%%%%%%%%%%%%%%%%%%%%%%%

\clearpage

%%%%%%%%%%%%%%%%%%%%%%%%%%%%%%%%%%%%%%%%%
\begin{table}
\caption{Collapse simulations. The simulations have been run with $N =
  10^5$ particles, except $S1^+$ for which $N = 8 \times 10^5$.  The
  columns list the initial virial ratio $u=(2K/|W|)_{t=0}$, the
  fractional mass loss $\Delta M = [M(t=0) - M(t_{end})]/M(t=0)$, the
  final global anisotropy $\kappa = 2K_r/K_T$ of the system of bound
  particles, its anisotropy radius $r_{\alpha}$ (i.e. the radius where
  the local anisotropy parameter is $\alpha=1$) in units of the
  half-mass radius, and the ellipticities $\epsilon=b/a$ and
  $\eta=c/a$, where $a \geqslant b \geqslant c$ are the axes of the
  final quasi-equilibrium state computed from the inertia
  tensor.}\label{tab:S}
\begin{center}
\begin{tabular}{ccccccc}
\hline \hline & $u$ &  $\Delta
M$ & $\kappa_{end}$ & $r_{\alpha}/r_M $ & $\epsilon $  & $\eta$  \\ \hline

$S1^+$ & 0.07 & 0.01 & 2.49 & 0.18 & 1.00 & 0.99 \\ %800K
$S1^*$ & 0.07 & 0.01 & 2.54 & 0.18 & 0.99 &0.98 \\%100K CGS
$S1^{\dagger}$ & 0.07 & 0.01 & 2.52 & 0.18 & 0.98 &0.97 \\%100K Falcon
$Sa$ & 0.05 & 0.03 & 2.75 & 0.18 & 0.98  & 0.97 \\  %S1 colder  100K
$Sb$ & 0.03 & 0.07 & 2.56 & 0.16 & 0.79 & 0.66 \\   %S1 colder unstable 100K
$S2$ & 0.08 & 0.10 & 2.16 & 0.30 & 1.00 & 0.98 \\
$S3$ & 0.25 & 0.00 & 2.13 & 0.29 & 0.99 & 0.98 \\

\hline

\label{tab:main}
\end{tabular}
\end{center}
\end{table}

\clearpage

%%%%%%%%%%%%%%%%%%%%%%%%%%%%%%%%%%%%%%%%%
\begin{table}
\caption{Simulation runs initialized with models constructed from the
  Jeans equations. The simulations have $2 \times 10^5$
  particles. Here we list the initial and final global anisotropy
  $\kappa = 2K_r/K_T$ of the system, its final anisotropy radius in
  terms of the half-mass radius, and the final ellipticities
  $\epsilon=b/a$ and $\eta=c/a$, where $a \geqslant b \geqslant c$ are
  the axes evaluated from the inertia tensor. The initial conditions
  are summarized in the last column and range from the best fit of
  $S1$, simulation $J1$, to regularized Jaffe profiles $\rho_{J}$ (see
  Eq.~\ref{eq:jaff}) with Osipkov-Merritt or $\f$ like
  anisotropy. Note that simulation $J4$, initialized with a
  regularized Jaffe density plus an unstable $\f$ anisotropy profile
  ($\kappa \approx 2.3$) evolves rapidly within the first dynamical
  time and the anisotropy is quickly reduced below $\kappa = 2$ while
  preserving the spherical symmetry.}\label{tab:jeans}
\begin{center}
\begin{tabular}{ccccccc}
\hline
\hline

 & $\kappa_{0}$ & $\kappa_{end}$ & $r_{\alpha}/r_M $ & $\epsilon $  & $\eta$ & notes:\\

J1 & 2.47 & 2.42 & 0.26 & 0.98 & 0.98 & \tiny{as S1} \\
J2 & 2.92 & 2.12 & 0.37 & 0.48 & 0.48 & \tiny{$\rho_J+\alpha_{S1}$}\\
J3 & 2.40 & 2.45 & 0.28 & 0.96 & 0.94 & \tiny{$\rho_J+\alpha_{OM}$}\\
J4 & 2.29 & 1.93 & 0.61 & 0.99 & 0.99 & \tiny{$\rho_J+\alpha_{\f}$}\\
J5 & 2.47 & 2.26 & 0.30 & 0.63 & 0.63 & \tiny{$\rho_{S1}+\alpha_{OM}$}\\

\hline

\end{tabular}
\end{center}
\end{table}

\clearpage

%%%%%%%%%%%%%%%%%%%%%%%%%%%%%%%%%%%%%%%%%

\end{document}